\documentclass[12pt,a4paper,leqno]{article}
\usepackage{graphicx}
\usepackage{color}
\title{Deep-water waves: On the nonlinear Schr{\"o}dinger equation and its solutions}
\author{Nikolay K. Vitanov$^1$\footnote{corresponding author, e-mail:
vitanov@imbm.bas.bg}, Amin Chabchoub$^2$, Norbert Hoffmann$^2$}
\date{$^1$ Institute of Mechanics, Bulgarian Academy of Sciences, Acad. G. Bonchev, Bl. 4, 1113 
Sofia, Bulgaria \\
$^2$ Institute of Mechanics and Ocean Engineering, Hamburg University of Technology, 21073 Hamburg, Germany}
\begin{document}
\maketitle
\begin{abstract}
We present a brief discussion on the nonlinear Schr{\"o}dinger equation
for modeling the propagation of the deep-water wavetrains and a discussion 
on its doubly-localized breather solutions that can be connected to the 
sudden formation of extreme waves, also known as rogue waves or freak waves.
\end{abstract}
\newpage
\section{Introduction}
Nonlinear phenomena are much studied in various area of science \cite{n1}-\cite{n6}.
Many of these phenomena are modeled by nonlinear partial differential equations or systems
of such equations, which increased the interest in the methods of obtaining exact and approximate
solutions of such equations \cite{a1}-\cite{a5}. In this brief report we shall be interested 
in extreme deep-water phenomena, also referred to as freak or rogue waves \cite{f1,f2} and its
description by breather solutions of the nonlinear Schr{\"o}dinger equation. Rogue water waves are 
extreme high sea waves that can cause severe damage on commercial and other ships or on oil platforms. 
Such waves are deep-water waves which probably can be described by breather solutions of 
equations that belong to the family of the nonlinear Schr{\"o}dinger equation. The latter will
determine the scope of our report as follows: (I) short notices on linear and nonlinear surface gravity water water; 
(II) Heuristic derivation the nonlinear Schr{\"o}dinger equation (NSE) for deep-water waves; 
(III) discussion on one extension of this equation (known as Dysthe equation); and (IV) a short discussion on breather 
solutions of the NSE obtained by Peregrine as well as Akhmediev and coworkers , which could 
be considered as appropriate models for the freak waves.
\section{Model equations for water waves: incompressible inviscid approximation}
\par 
We start from the differential form the basic equation of fluid mechanics are as follows \cite{c4}
\begin{eqnarray}\label{a1}
\frac{\partial \rho}{\partial t} + \nabla \cdot (\rho \vec{v}) &=& 0 \ : \
{\rm  mass \ conservation}, \nonumber \\
\frac{\partial \vec{v}}{\partial t} + (\vec{v} \cdot \nabla) \vec{v} &=&
- \frac{1}{\rho} \nabla p + \vec{f} + \frac{\eta^*}{\rho} \nabla^2 \vec{v} \ :
{\rm momentum \ conservation}.
\end{eqnarray}
In Eqs(\ref{a1}) $\rho$ is the density of the fluid; $\vec{v}$ is the fluid velocity;
$p$ is the pressure; $\vec{f}$ summarizes the body forces acting on the fluid;
$\eta^*$ is coefficient of viscosity called kinematic viscosity.
\par
In addition boundary conditions should be imposed. We assume that the depth of the
fluid is $h$ and it is bounded from below by a hard horizontal bed. The upper
fluid surface is assumed to be free. The unperturbed free upper surface is at $z=0$. When the upper
surface is perturbed, there is vertical displacement $\eta(x,y,t)$ of each point
of the surface. Then, the boundary condition on the upper fluid surface is at $z=\eta(x,y,t)$.
On the lower solid surface the normal component of velocity has to vanish, i.e. no flux is permitted at the bottom. 
That is, $v_z=0$ at $z=-h$.
\par
We shall discuss the case of constant fluid density $\rho = 
{\rm const}$ (incompressible fluid approximation). In addition, the waves will have wavelength much longer than
approximately $1.8$ cm. For this case  the viscosity effects are negligible, i.e., we shall
consider gravity waves (and not capillary waves).
\subsection{Irrotational approximation and model of small amplitude 
surface gravity waves}
The quantity
$\vec{\omega} = \nabla \times \vec{v}$ is called vorticity of the
flow and when $\vec{\omega}=0$ the flow is called irrotational. In this case the
velocity of the flow is a potential field: $\vec{v} = \nabla \phi$, where
$\phi$ is the velocity potential (the central quantity we shall discuss below).
In addition, for gravity waves $\vec{f} = -g \vec{e}_z$, where $g$ is the
acceleration of gravity and $\vec{e}_z = (0,0,1)$. For small water surface amplitudes and 
in the irrotational approximation, the model equations (\ref{a1})
and the boundary conditions are reduced as follows 
\begin{eqnarray}\label{b1}
\nabla^2 \phi &=& 0; \ -h < z < \eta (x,y,t) \to \ {\rm from \ mass \ conservation}, 
\nonumber \\
\frac{\partial \phi}{\partial t} &=& -\left[ \frac{1}{2} \left( 
\left(\frac{\partial \phi}{\partial x} \right)^2 + 
\left( \frac{\partial \phi}{\partial y} \right)^2+ 
\left( \frac{\partial \phi}{\partial z} \right)^2\right)+ \eta g \right]; \
{\rm at} \ z = \eta(x,y,t) \to \ {\rm from } \nonumber \\
&& {\rm  \ momentum \ conservation}, \nonumber \\
\frac{\partial \phi}{\partial z} &=& \frac{\partial \eta}{\partial x}
\frac{\partial \phi}{\partial x} + \frac{\partial \eta}{\partial y}
\frac{\partial \phi}{\partial y} + \frac{\partial \eta}{\partial z}
\frac{\partial \phi}{\partial z}; \ {\rm at} \ z = \eta(x,y,t) \to \ {\rm from \ b.c. \ on } \nonumber \\
&& {\rm \ the \ top \ surface}, \nonumber \\
\frac{\partial \phi}{\partial z} &=& 0; \ {\rm at} \ z=-h\to \ {\rm from \ b.c. \ on 
\ the \ bottom \ surface}.
\nonumber\\
\end{eqnarray}
Thus, the model equation becomes linear (the Laplace equation) but the boundary conditions are nonlinear.
\subsection{Shallow-water waves and deep-water waves}
For small amplitudes (but long wavelengths) water waves, the nonlinear relationships from the system (\ref{b1}) 
can be linearized. If the
mean surface displacement and the mean velocity potential are small with respect to the wavelength and 
to wave period scales, then, the nonlinear terms in the boundary conditions can be neglected. 
After a Taylor series expansion of the small quantity $\eta$ 
(and keeping only the first term of the expansion) the top boundary condition can be written as condition on $z=0$.
Thus, we obtain the following simplified (and linear) problem
 \begin{eqnarray}\label{c1}
\nabla^2 \phi &=& 0; \ -h < z < 0,
\nonumber \\
\frac{\partial^² \phi}{\partial t²} &=& - g \frac{\partial \phi}{\partial z} ; \
{\rm at} \ z = 0,  \nonumber \\
\frac{\partial \phi}{\partial z} &=& 0; \ {\rm at} \ z=-h.
\end{eqnarray}
The next approximation is that we assume that the waves propagate in the $x$-direction and are uniform in the $y$-direction.
Thus, the problem becomes one-dimensional and one searches for a traveling wave solution with wave frequency $\omega$
and wavenumber $k$:
\begin{equation}\label{c2}
\phi(x,t) = \overline{A}(x,z) \sin(k x - \omega t).
\end{equation}
The substitution of Eq. (\ref{c2}) in (\ref{c1}) leads to the following solutions for the velocity potential $\phi$
and for the surface displacement $\eta$
\begin{eqnarray}\label{c3}
\eta = A \cos(kx - \omega t); \ \phi = \omega A \frac{\cosh (k(z+h))}{k \sinh(kh)} 
\sin(kx - \omega t); \nonumber \\
A = 2 \frac{a k }{\omega} \exp(-k h) \sinh(k h),
\end{eqnarray}
where $a$ is a constant of integration, $A$, $k$ and $\omega$ denote the wave amplitude, the wavenumber 
and the wave frequency, respectively. The dispersion relation for the small amplitude surface water waves as well as their phase
velocity $v$ and group velocity $v_g$ are as follows
\begin{eqnarray}\label{c4}
\omega^2 = gk \tanh(kh); \ v = \sqrt{\frac{g}{k} \tanh(kh)}; 
\ v_g = \frac{v}{2} \left[ 1 + \frac{2kh}{2 \sinh(2 kh)}\right].
\end{eqnarray}
The relationship $R=\frac{\rm depth}{\rm wavelength}=\frac{h }{\lambda}$ has two limit cases:
$R<<1$ (shallow-water waves) and $R>>1$ (deep-water waves). For the case of shallow-water waves the
dispersion relation (\ref{c4}) can be approximated by
$$
\omega = k \sqrt{gh} \left[ 1 - \frac{k^2 h^2}{6}+ \dots \right]; \ c_0 = \sqrt{g h}.
$$
For very long shallow water waves $\omega = k c_0$; $v=\frac{\omega}{k}\approx c_0$;
$v_g = \frac{\partial \omega}{\partial k} \approx k_0$. For deep-water waves the approximation for the dispersion 
relation is $\omega \approx \sqrt{g k}$ and the phase and group velocity are $v = \sqrt{g/k}$; $v_g = \sqrt{g/(2 k)}$, respectively. 
That is, the group velocity is smaller and half the phase velocity.
\section{Deep-water waves. The nonlinear Schr{\"o}dinger equation. The Dysthe equation}
A weakly nonlinear model for shallow-water waves can be described by the
Korteweg-de Vries equation. We shall be interested in the deep-water case, which can 
be modeled by the nonlinear Schr{\"o}dinger equation.
\subsection{Derivation of the nonlinear Schr{\"o}dinger equation by applying a Taylor series 
expansion to the dispersion relation  for deep-water waves}
A weakly nonlinear approximation to the nonlinear deep-water wave problem are the Stokes waves. They however are unstable against
modulation perturbations. The velocity of the Stokes waves to second-order in steepness is
\begin{equation}\label{c5}
v=\sqrt{\frac{g}{k}(1+\frac{k^2 a^2}{2})},
\end{equation}
where $a$ denotes now the wave amplitude. From Eq. (\ref{c5}) one obtains easily the dispersion relation
$\omega = \sqrt{gk(1+\frac{k^2 a^2}{2})}$. Let us consider a slowly modulated  Stokes wave wavetrain
\begin{equation}\label{d0}
\eta = Re[A(X,T) \exp(i(\omega_0 t - k_0 x))],
\end{equation}
 where $\omega_0$ and $k_0$ are the frequency and
wave number of carrier Stokes wave and $A(X,T)$ is the modulation amplitude of the wavetrain. In
addition, $X = \epsilon x$ and $T=\epsilon t$ ( $\epsilon <<1$) are the slowly varying space and time 
variables, respectively. Physically, $\varepsilon:=Ak_0$ is the steepness of the wave and is assumed to be small. 
Let us now perform a Taylor series expansion
around the wavenumber  $k_0$ and the amplitude $A_0=A(0,0)$...
 The dispersion relation of the carrier Stokes wave is
\begin{equation}\label{d01}
\omega = \sqrt{g k (1+ k^2 \mid A \mid^2)},
\end{equation}
where $\mid A \mid$ is  the amplitude of the Stokes wave (and the amplitude of the
envelope). 
The Taylor series expansion about the wavenumber $k_0$ of the
carrier wave and about the envelope amplitude $A=A_0=0$ as follows \cite{c4}
\begin{equation}\label{d1}
\omega = \omega_0 + \frac{\partial \omega}{\partial k}(k-k_0) +
\frac{1}{2} \frac{\partial^2 \omega}{\partial k^2}(k-k_0)^2 +
\frac{\partial \omega}{\partial \mid A \mid^2}(\mid A \mid^2-\mid A_0 \mid^2).
\end{equation} 
Let $\Omega = \omega - \omega_0$ and $K=k-k_0$. In addition (accounting also for Eq.(\ref{d01})) 
$\frac{\partial \omega}{\partial k}\mid_{k=k_0} = v_g = \frac{\omega_0}{2 k_0}$ (note that
the group velocity of the envelope is twice smaller than the phase velocity of the 
carrier wave); $\frac{\partial^2 \omega}{\partial k^2}\mid_{k=k_0} = 2P = -\frac{\omega_0}{8 k_0^2}$; 
$Q=\frac{\partial \omega}{\partial \mid A \mid^2}\mid_{A_0=0}= \frac{1}{2} \omega_0 k_0^2$. 
Then, from Eq.(\ref{d1})
\begin{equation}\label{d2}
\Omega =   v_g K + P K^2 + Q \mid A \mid^2.
\end{equation}
The Fourier and the inverse Fourier transform of the envelope function are
\begin{eqnarray}\label{d3}
A(K,\Omega) = {\cal{F}}[A(X,T)] = \int_{-\infty}^{\infty} dXdT\  A(X,T) \
\exp[i(\Omega T - KX)], \nonumber \\
A(X,T) = {\cal{F}}^{-1}[A(K,\Omega)] = \left( \frac{1}{2 \pi} \right)^2 
\int_{-\infty}^{\infty} dK d\Omega \ A(K,\Omega) \
\exp[-i(\Omega T - KX)]. \nonumber \\ 
\end{eqnarray}
From Eqs. (\ref{d3})
\begin{equation}\label{d5}
\frac{\partial A}{\partial X} = iK {\cal F}^{-1} [A(K, \Omega)], \
\frac{\partial A}{\partial t} = -i \Omega {\cal F}^{-1} [A(K, \Omega)].
\end{equation}
$\Omega$ and $K$ are of  order $\epsilon$. Then from Eq.(\ref{d5}) we can
write
\begin{equation}\label{d6}
K = - i \epsilon \frac{\partial}{\partial X}; \
\Omega = i \epsilon \frac{\partial}{\partial T}.
\end{equation}
The  substitution of the relationships from Eq.(\ref{d6}) in Eq.(\ref{d2}) and application
of the resulting operator equation to the envelope amplitude $A$ leads to the
nonlinear Schr{\"o}dinger equation for the evolution of the amplitude of the envelope
of the wavetrain ($\epsilon$ is incorporated in $T$ and $X$ by appropriate rescaling).
\begin{equation}\label{d7}
i \left( \frac{\partial A}{\partial T} + \frac{\omega_0}{2 k_0} \frac{\partial A}{\partial X} 
\right) - \frac{\omega_0}{8 k_0^2}\frac{\partial^2 A}{\partial X^2} - \frac{1}{2} \omega_0 k_0^2
\mid A \mid^2 A =0.
\end{equation}
Eq. (\ref{d7}) can be rescalled as follows: $\tau=-\frac{\omega_0}{8 k_0^2} T$; $\xi = X-v_gT =
X-\frac{\omega_0}{2 k_0} T$ (coordinate is in a frame that moves with the group velocity of the
wavetrain); $q=\sqrt{2}k_0^2 A$. The rescalled form of the nonlinear Schr{\"o}dinger equation is
\begin{equation}\label{d8}
i \frac{\partial q}{\partial \tau} + \frac{\partial^2 q}{\partial \xi^2} + 2 \mid q \mid^2 q =0.
\end{equation}
\subsection{More accurate deep-water wave envelope equation: The Dysthe
equation}
The Dysthe equation \cite{c5} is an extension of the nonlinear Schr{\"o}dinger equation. This equation
aims to solve the problem with the bandwidth limitation of the NSE.
The nonlinear Sch{\"o}rdinger equation is valid for small steepness values. i.e. $k_0A <<1$ and when the bandwidth is narrow
($\Delta k / k <<1$, $\Delta k$ is the modulation wavenumber). In order to obtain the Dysthe equation
one starts with the model equations for the velocity potential $\phi(x,y,z,t)$ and surface displacement
$\eta(x,y,t)$ for an incompressible inviscid fluid with uniform depth $h$
\begin{eqnarray}\label{e1}
\nabla^2 \phi = 0 , \ {\rm for} - h < z< \eta, \nonumber \\
\frac{\partial^2 \phi}{\partial t^2} + g \frac{\partial \phi}{\partial z} + \frac{\partial}{\partial t}(\nabla \phi)^2
+ \frac{1}{2} \nabla \phi \cdot \nabla (\nabla \phi)^2 = 0, \ {\rm at} z = \eta, \nonumber \\
\frac{\partial h}{\partial t} + \nabla \phi \cdot \nabla h = \frac{\partial \phi}{\partial z}, \
{\rm } z=h, \nonumber \\
\frac{\partial \phi}{\partial z} = 0, \ {\rm at } z=-h. 
\end{eqnarray}
Next one assumes that $k A = O(\epsilon)$; $\frac{\Delta k}{k} = O(\epsilon)^{1/2}$, $(k h)^{-1} = O (\epsilon^{1/2})$
and performs the following expansion for the velocity potential $\phi$ and surface displacement $h$
\begin{eqnarray}\label{e2}
\phi = \overline{\phi} + \frac{1}{2} \left(A \exp(i \theta + kz) + A_2 \exp(2(i \theta + kz)) + \dots
+ {\rm c.c.} \right),
\nonumber \\
h = \overline{h} + \frac{1}{2} \left(B \exp(i \theta + kz) + B_2 \exp(2(i \theta + kz)) + \dots +
{\rm c.c.} \right),
\nonumber \\
\end{eqnarray}
where $\theta = kx - \omega t$. The drift $\overline{\phi}$, the set down $\overline{h}$ and the amplitudes
$A,A_2,\dots,B,B_2,\dots$ are functions of the slow modulation variables $\epsilon x, \epsilon y, \epsilon t$.
Furthermore, a non-dimensionalization of the variables is performed as follows: $\omega t \to t$; $k(x,y,z) \to (x,y,z)$;
$k^2 \omega^{-1} (A, \dots, A_n, \overline{\phi}) \to (A, \dots, A_n, \overline{\phi})$; 
$k(B,\dots,B_n,\overline{\eta})\to (B,\dots,B_n,\overline{\eta})$. Note that the functions $\overline{\phi}$; $\overline{\eta}$;
$A,\dots, A_n$; $B, \dots, B_n$ are functions of the variables $\epsilon^{1/2}(x,z,t)$. Proceeding with
the expansions up to order $O(\epsilon^{3.5})$, one obtains the Dysthe equation
\begin{eqnarray}\label{e3}
\frac{\partial A}{\partial t} + \frac{1}{2}\frac{\partial A}{\partial x} + \frac{i}{8} \frac{\partial^2 A}{\partial x^2} -
\frac{i}{4} \frac{\partial^2 A}{\partial y^2} - \frac{1}{16} \frac{\partial^2 A}{\partial x^3} + \frac{3}{8} \frac{\partial^3 A}{\partial x \partial y^2}
- \frac{5i}{128} \frac{\partial^4 A }{\partial x^4} + \nonumber \\
\frac{15i}{32} \frac{\partial^4 A }{\partial x^2 \partial y^2} - \frac{3i}{32} \frac{\partial^4 A }{\partial y^4} + \frac{i}{2}
\mid A \mid^2 A + \frac{7}{256} \frac{\partial^5 A }{\partial x^5} - \frac{35}{64} \frac{\partial^5 A }{\partial x^3 \partial y^2} +
\frac{21}{64} \frac{\partial^5 A }{\partial x \partial y^4} + \nonumber \\
\frac{3}{2} \mid A \mid^2 \frac{\partial A}{\partial x} - \frac{1}{4} A^2 \frac{\partial A^*}{\partial x} +
i A \frac{\partial \overline{\phi}}{\partial x} = 0, 
\end{eqnarray}
for the corresponding boundary conditions
\begin{eqnarray}\label{e4}
\nabla^2 \overline{\phi} = 0, \ {\rm for} \ -h < z < 0, \nonumber \\
\frac{\partial \overline{\phi}}{\partial z} = \frac{1}{2} \frac{\partial \mid A \mid^2}{\partial x}, \ {\rm at} 
\ z=0, \nonumber \\
\frac{\partial \overline{\phi}}{\partial z} = 0, \ {\rm at} \ z= -h.
\end{eqnarray}
\section{Peregrine and Akhmediev-Peregrine breathers}
Akhmediev and co-workers \cite{c2, c3} derived a family of space-periodic pulsating solutions of the NSE, 
also called Akhemdiev breathers, that start from the plane wave solution at $\tau = -\infty$ and return again to
the plane wave form at $\tau = \infty$. Taking this period to infinity, we obtain the Peregrine 
breather \cite{c1}, a solution which is doubly-localized, in space and time, and pulsates only once. 
These properties make the Peregrine breather an ideal model to describe oceanic rogue waves. In the 
same work \cite{c2}, Akhmediev and co-workers derived a higher-order doubly-localized breathers, 
also referred to the Akhmediev-Peregrine breather. In fact, there is an infinite hierarchy of 
doubly-localized Akhmediev-Peregrine solutions. Generally, the $j-$th Akhmediev-Peregrine solution can be written in terms of polynomials:
\begin{equation}\label{d9}
q_j(\xi,\tau) = q_0 \exp(2 i \mid q_0 \mid^2 \tau) \left[ (-1)^j+ \frac{G_j+i H_j}{D_j} \right],
\end{equation}
where $q_0$ is proportional to the carrier wave amplitude and $G_j(\xi,\tau)$; $H_j(\xi,\tau)$
and $D_j(\xi,\tau)$ are appropriate polynomials. The first-order rational solution (Peregrine breather) is given by
\begin{equation}\label{d10}
G_1 = 4; \ H_1 = 16 \mid q_0 \mid^2 \tau; \ D_1 = 1 + 4 \mid q_0 \mid^2 \xi^2 + 16 \mid q_0 \mid^4
\tau^2
\end{equation}
The second\textcolor{red}{-}order rational solution (Akhmediev-Peregrine breather) is given by
\begin{eqnarray}\label{d11}
G_2 = \left( \mid q \mid^2 \xi^2 + 4 \mid q_0 \mid^4 \tau^2 + \frac{3}{4} \right)
\left( \mid q_0 \mid^2 \xi^2 + 20 \mid q_0 \mid^4 \tau^4 + \frac{3}{4} \right) - \frac{3}{4} 
\nonumber \\
H_2 = 2 \mid q_0 \mid^2 \tau (4 \mid q_0 \mid^4 \tau^2 - 3 \mid q_0 \mid^2 \xi^2) +
2 \mid q_0 \mid^2 \tau \bigg( (2 \mid q_0 \mid^2 \xi^2 + \nonumber \\ 
4 \mid q_0 \mid^4 \tau^2)^2 - \frac{15}{8} \bigg ) \nonumber \\
D_2=\frac{1}{3} (\mid q_0 \mid^2 \xi^2 + 4 \mid q_0 \mid^4 \tau^2)^3 + 
\frac{1}{4} (\mid q_0 \mid^2 \xi^2 - 12 \mid q_0 \mid^4 \tau^2)^2 + \nonumber \\
\frac{3}{64} (12 \mid q_0 \mid^2 \xi^2 + 176 \mid q_0 \mid^4 \tau^2 +1)
\end{eqnarray} 
Figure 1 shows the two solutions. 
\begin{figure}[h]
\centering
\includegraphics[width=9cm]{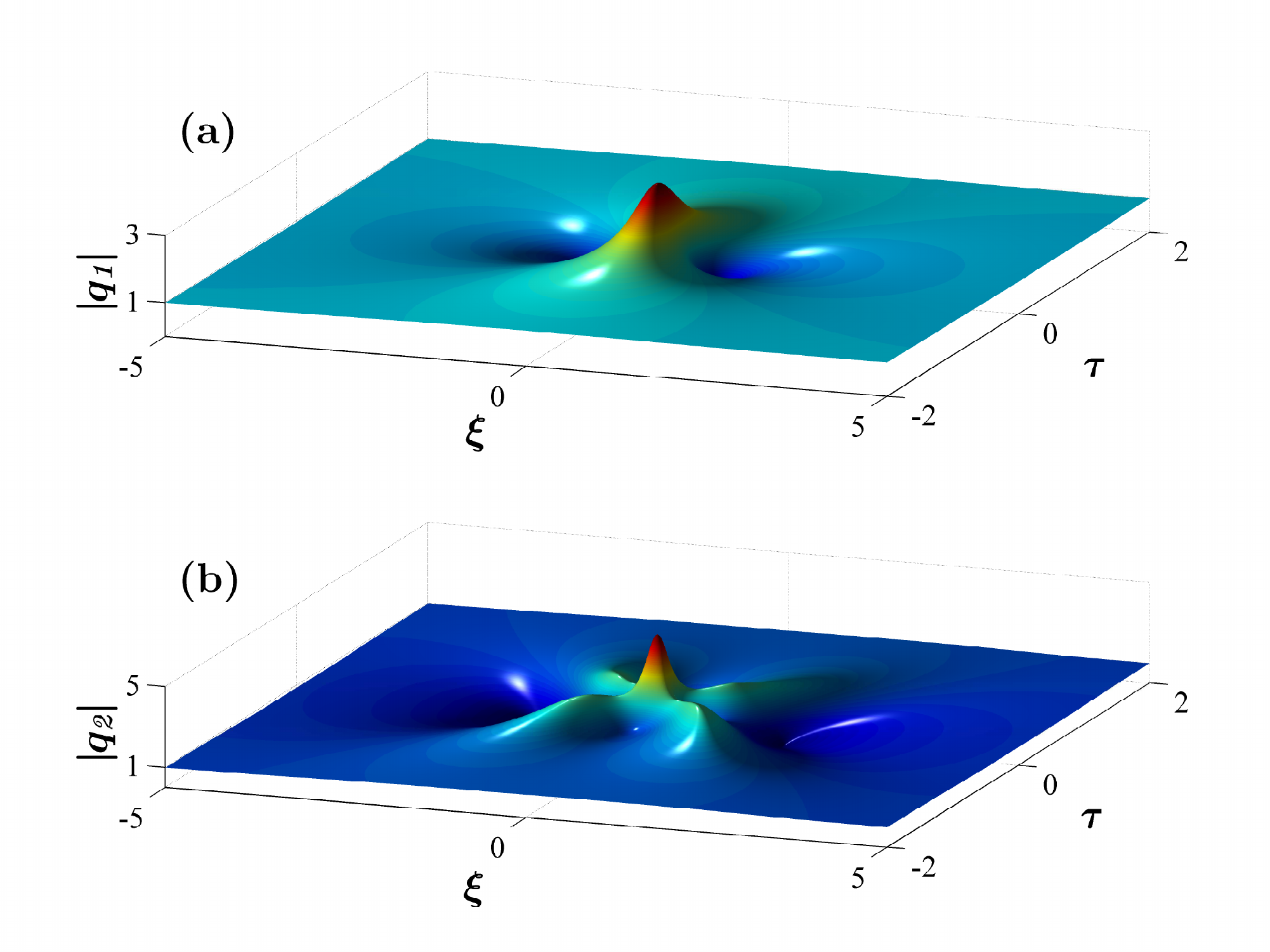}
\caption{(a) First-order rational solution (Peregrine breather), which amplifies the carrier-amplitude by a factor of three. (b) Second-order rational solution (Akhmediev-Peregrine breather), which amplifies the carrier-amplitude by a factor of five. Both solutions are localized in both, space and time.}\label{fig1}
\end{figure}
The Peregrine and Akhmediev-Peregrine breathers up to fifth-order have been recently observed in a deep-water
waves flume \cite{c6}-\cite{c9}. We intend to conduct in the future theoretical, numerical and 
experimental investigations of NSE solutions as well as other equations of this kind,
based on the method of simplest equation \cite{b1}-\cite{b3} and similar.


\begin{thebibliography}{99}
\bibitem{n1}
SCOTT, A. C. Nonlinear  Science. Emergence and Dynamics of
	Coherent Structures. Oxford, UK, Oxford University Press, 1999.
\bibitem{n2}
PANCHEV S., T. SPASSOVA, N.K. VITANOV. Analytical and numerical investigation of two 
	families of Lorenz-like dynamical systems. {\sl Chaos, Solitons \& 
	Fractals} {\bf 33} (2007) 1658-1671.
\bibitem{n3}
KANTZ H., D. HOLSTEIN, M. RAGWITZ, N. K. VITANOV. Markov Chain Model for Turbulent
	Wind Speed Data. {\sl Physica A} {\bf 342} (2004) 315 - 321.
\bibitem{n4}
BOECK T., N. K. VITANOV. Low-Dimensional Chaos in Zero-Prandtl-Number 
	Benard-Marangoni Convection. {\sl Phys. Rev. E} {\bf 65} (2002) Article No:037203.

\bibitem{n5}
VITANOV N. K, I. P. JORDANOV, Z. I. DIMITROVA On Nonlinear Population Waves,
	{\sl Applied Mathematics and Computation} {\bf 215} (2009) 2950-2964.
\bibitem{n6}
VITANOV N. K.  Upper Bounds on the Heat Transport in a Porous Layer.  {\sl Physica D}
	{\bf 136} (2000) 322 - 339. 
\bibitem{a1}
AKHMEDIEV N. N., A. ANKIEWICZ. Solitons. Nonlinear Pulses and Beams. London, Chapman \& Hall,
	1997.
\bibitem{a2}
HIROTA, R. Exact Solution of Korteweg-de Vries Equation for
	Multiple Collisions of Solitons. {\sl Phys. Rev. Lett.},  {\bf 27} (1971) 1192 - 1194.
\bibitem{a3}
KUDRYASHOV, N. A., M. B. SOUKHAREV. Popular Ansatz Methods and Solitary Wave Solutions of
	 the Kuramoto - Sivashinsky Equation. {\sl Regular \& Chaotic Dynamics} {\bf 14} (2009)
	  407 - 419.
\bibitem{a4}
VITANOV N. K. Modified Method of Simplest Equation: Powerful Tool for Obtaining Exact and 
	Approximate Traveling-Wave Solutions of Nonlinear PDEs. {\sl Commun. Nonlinear Sci.  Numer. 
	Simulat.}  {\bf 16} (2011) 1176 - 1185.
\bibitem{a5}
VITANOV, N. K., Z. I. DIMITROVA, H. KANTZ Modified method of simplest equation
	and its application to nonlinear PDEs. {\sl Applied Mathematics and Computation},
	{\bf 216} (2010) 2587-2595.
\bibitem{f1}
KHARIF C., E. PELINOVSKY, A. SLUNYAEV. Rogue waves in the ocean. Berlin, Sprinfer, 2009.
\bibitem{f2}
PELINOVSKY E., C. KHARIF. (Eds.) Extreme ocean waves. Berlin, Springer, 2008.  
\bibitem{c2}
AKHMEDIEV N., V. M. ELEONSKII,  N. E. KULAGIN. Generation of of a periodic sequence 
           of picosecond pulses in an optical fiber. Exact solutions. 
	   {\sl Sov. Phys. JETP} {\bf 89} (1985)  1542-1551.
\bibitem{c3}
AKHMEDIEV N.,  V. I. KORNEEV. Modulation instability and periodic solutions 
          of the nonlinear Schr{\"o}dinger equation. {\sl  Theor. Math. Phys. (USSR)}  
	  {\bf 69} (1987) 1089-1093. 
\bibitem{c1}
PEREGRINE, D. H. Water waves, nonlinear Schrödinger equations and their solutions. 
           {\sl J. Austral. Math. Soc. B} {\bf 25} (1983) 16-43.
\bibitem{c4}
REMOISSENET M.  Waves Called Solitons. Berlin, Springer, 1993.
\bibitem{c5}
DYSTHE K. B. Note on a modification to the nonlinear Schr{\"o}dinger equation for
	application to deep water waves. {\sl Proc. Roy. Soc. London, Ser. A} 
	{\bf 369} (1979), 105-114.
\bibitem{c6}
CHABCHOUB A., N. VITANOV, N. HOFFMANN. Experimental evidence for breather type 
  	dynamics in freak waves. {\sl PAMM}  {\bf 10}, No. 1 (2010) 495-496.
\bibitem{c7}
CHABCHOUB A., N. P. NOFFMANN, N. AKHMEDIEV N. Rogue wave observation in a water wave tank.
{\sl Physical Review Letters} {\bf 106}, No. 20, (2011), Article No. 204502
\bibitem{c8}
CHABCHOUB A., N. P. NOFFMANN, M. ONORATO, N. AKHMEDIEV.	
	Super Rogue Waves: Observation of a Higher-Order Breather in Water Waves.
	{\sl Physical Review X} {\bf 2}  No.1 (2012), Article No. 011015.
\bibitem{c9}
CHABCHOUB A., N. HOFFMANN, M. ONORATO, A. SLUNYAEV, A. SERGEEVA, E. PELINOVSKY, N. AKHMEDIEV. Observation of a hierarchy of up to fifth-order rogue waves in a water tank. {\sl Physical Review E} {\bf 86}  No.1 (2012), Article No. 056601. 
\bibitem{b1}
VITANOV N. K., Z. I. DIMITROVA.  Application of the Method of Simplest Equation for 
	Obtaining Exact Traveling-Wave Solutions for Two Classes of Model PDEs From Ecology
	and Population Dynamics. {\sl Commun. Nonlinear Sci.  Numer. Simulat.} 
	{\bf 15} (2010) 2836-2845.
\bibitem{b2}
VITANOV N. K. Application of Simplest Equations of Bernoulli and Riccati kind
	for Obtaining Exact Traveling Wave Solutions for a Class of PDEs with
	Polynomial Nonlinearity. {\sl Commun. Nonlinear Sci.  Numer. Simulat.} 
	{\bf 15} (2010)  2050 - 2060.
\bibitem{b3}
VITANOV N. K., Z. I. DIMITROVA, N. K. VITANOV.  On The Class of Nonlinear PDEs that can
	be Treated by the Modified Method of Simplest Equation. Application to Generalized
	Degasperis-Processi Equation and b-Equation. {\sl Commun. Nonlinear
	Sci. Numer. Simulat.}  {\bf 16} (2011) 3033 - 3044.
\end{thebibliography}
\end{document}